\documentclass[%
 reprint,
 amsmath,amssymb,
 aps,
]{revtex4-2}
\usepackage{url}
\usepackage{xcolor}
\usepackage{enumitem}
\usepackage{graphicx}
\usepackage{dcolumn}
\usepackage{bm}
\usepackage{natbib}
\usepackage{caption}
\usepackage{subcaption}
\usepackage{siunitx}
\usepackage{hyperref}
\usepackage{natbib}
\hypersetup{
    colorlinks=true,
    linkcolor=blue,
    filecolor=magenta,      
    urlcolor=blue,
    pdftitle={Overleaf Example},
    pdfpagemode=FullScreen,
    }

\begin{document}

\preprint{APS/123-QED}

\title{Cosmic filaments confirm unexplained CMB temperature decrements in two independent redshift ranges}


\author{Juan Ignacio Domínguez Feldman}
 \email{juan.dominguez99@mi.unc.edu.ar}
\affiliation{Facultad de Matemática, Astronomía, Física y Computación (FAMAF), UNC, Córdoba, Argentina}
\author{Luis A. Pereyra}%
 \email{luis.pereyra@unc.edu.ar}
 \affiliation{Observatorio Astronómico de Córdoba (OAC), UNC, Córdoba, Argentina}
\author{Frode K. Hansen}
\email{f.k.hansen@astro.uio.no}
\affiliation{Institute of Theoretical Astrophysics, University of Oslo, PO Box 1029 Blindern, 0315 Oslo, Norway}
\author{Facundo Toscano}
\email{facundo.toscano@mi.unc.edu.ar}
\affiliation{Instituto de Astronomía Teórica y Experimental (IATE), CONICET-UNC, Córdoba, Argentina}
\author{Diego Garcia Lambas}
\email{diego.garcia.lambas@unc.edu.ar}
\affiliation{Instituto de Astronomía Teórica y Experimental (IATE), CONICET-UNC, Córdoba, Argentina\\
Observatorio Astronómico de Córdoba (OAC), UNC, Córdoba, Argentina}

\date{\today}

\begin{abstract}
  Recent papers have reported an unexplained cooling of CMB photons passing through galaxies in nearby cosmic filaments $z<0.02$ at the $>5\sigma$ level. Here we show for the first time that this effect is also present at higher redshifts $0.02<z<0.04$. Instead of calculating the CMB temperature around individual galaxies as in previous works, we analyze mean CMB temperature profiles associated to cosmic filaments in three dimensions. We have considered different thresholds in the linear K-band luminosity density of the filaments as a proxy to mass density. Furthermore, we have analyzed the dependence of the results on the average orientation of filaments with respect to the line of sight. These studies were implemented to test the expected dependence on mass density as well as on photon trajectory length within the cosmic filaments. We find a $3-4\sigma$ detection of a CMB temperature decrement trend towards the spine of the filaments, the larger the mass and the more radially oriented the filament, the stronger the CMB temperature decrement. This trend is seen independently in both redshift ranges $0.004<z<0.02$ and $0.02<z<0.04$. We therefore conclude that our results provide strong evidence for a lower CMB temperature along massive cosmic filaments in the nearby universe $z<0.04$.
\end{abstract}
\maketitle

\section{Introduction}
The prediction and subsequent discovery of the Cosmic Microwave Background (CMB) stands as one of the greatest breakthroughs of modern cosmology. Since then, detailed studies of the CMB have played a key role in the development of theoretical models of the universe. This remnant radiation from the early universe is remarkably homogeneous\cite{COBE},  exhibiting only small-amplitude fluctuations \cite{10.23943/princeton/9780691209838.001.0001}, which are fundamental for understanding the formation of present-day large-scale structures. \\

Among these structures are cosmic filaments, corresponding to elongated overdense regions tracing the highly anisotropic distribution of galaxies. These filaments can extend over scales larger than $100$ Mpc in length \cite{Bond1996}, while their typical width along the spine is only a few Mpc \citep{Cautun2014}. In large-scale numerical simulations, filaments grow by gravitational instability from sparse coherent inhomogeneities in the expanding substratum \citep{GalarragaEspinosa2024, Wang2024}. Within this framework, the present-day filaments of galaxies can be interpreted as direct evidence of significant gravitational compression perpendicular to the filament spine \citep{Kraljic2018, Laigle2018}. \\

From the perspective of the matter content of the universe, filaments constitute a substantial fraction of both baryonic and dark matter in the Lambda Cold Dark Matter model ($\Lambda$CDM model). Recent analyses of numerical simulations \cite{GalarragaEspinosa2024,Gal_rraga_Espinosa_2022} have further revealed distinct behaviors between short and long filaments: short filaments tend to collapse along their spine, while long filaments exhibit high expansion velocities along their spine axis. \\
    
In our previous papers (Luparello et al. 2023 (L2023), Hansen et al. 2023 and 2025 (H2023 and H2025), Garcia Lambas et al. 2024 \cite{Luparello_2022, Hansen_2023, Hansen2024,garcialambas2024}) we have shown that the observed CMB temperature around large, nearby spiral galaxies within $z<0.02$ is significantly ($20-30\mu$K) lower than expected, as confirmed comparing to $10.000$ Planck-like synthetic CMB simulations. The significance of this effect exceeds $4\, \sigma$. As noted in those works, the observed CMB temperature decrement extends well beyond the virial radius of the galaxies, suggesting that large-scale inhomogeneities play an important role in this newly identified extragalactic systematic. In H2025, it was further shown that this effect becomes even more prominent in regions associated with the densest filaments-approximately half of the large spirals in the sample lie in such environments. When the largest angular scale modes of the CMB are removed from the analysis, reducing noise from intrinsic CMB fluctuations, the significance of the CMB temperature profile go to nearly $6 \, \sigma$. The effect has also been confirmed by cross-correlating the nearby dark matter distribution with the CMB temperature field in \citep{Cruz}.\\

Despite this, Toscano et al. 2025 \cite{Toscano2025} showed that masking regions around these galaxies does not significantly affect the estimation of cosmological parameters within the $\Lambda$CDM framework. Additionally, Hansen et al 2025b \cite{Hansen2025b} found that CMB photons passing through local voids ($z<0.03$) are, contrary to expectation, systematically hotter (not colder) than the mean CMB temperature, with a significance of $2.7 - 3.6 \sigma$.  Combined with the observations in this as well as our previous papers, this is potentially pointing to a sign change of the Integrated Sachs-Wolfe (ISW) effect in the local universe. \\

Taking these findings into account, in this work we investigate the connection between the previously reported CMB temperature decrement around overdense regions and the presence of filaments in the nearby universe. In particular, we will extend the redshift range beyond $z<0.02$ were the effect has been analysed and detected so far. We focus particularly on the densest filaments and those oriented along the line of sight, where the signal is expected to be strongest due to the longer trajectory length of the photons in these filaments. We will first attempt to confirm previous findings at the lowest redshifts $z<0.02$ using our new approach based on CMB temperature profiles around filaments instead of independent galaxies, and then check if a similar signal may be present in the next redshift shell $0.02<z<0.04$.\\

Tanimura et al. 2020 \cite{Tanimura2020} investigated the hot gas associated with filaments identified in the Sloan Digital Sky Survey (SDSS) \cite{Kollmeier2019}, within the redshift range $0.2 < z < 0.6$. They detected thermal Sunyaev-Zel'dovich (tSZ) signal in these structures with a significance of $4.4\, \sigma$, estimating the gas temperature to be around $10^6\, K$. The tSZ effect lowers the temperature of CMB photons scattered on this gas for photon frequencies below $217$Ghz. However, as was shown in H2025, this strongly frequency dependent effect can account only for a maximum $\sim2\mu$K CMB temperature decrement for the frequencies below 217GHz which we analyse, much less than the $20-30\mu$K frequency independent decrement observed in galaxies. Here, we will show that for our filaments, the associated temperature decrements due to the tSZ effect are also less than $2\mu$K.

The paper is organized as follows: In Section \ref{Data} we present the data used in this work, including the filament identification procedure and the CMB maps. Section \ref{Methodology} describes our methodology, including the computation of perpendicular distances to filament spines, the removal  of large-scale multipoles and the calculation of CMB temperature profiles associated with the filaments. In Section \ref{Results}, we present our main results, focusing on the dependence of the CMB temperature profiles on filament luminosity density and orientation relative to the line of sight. Finally, Section \ref{Discussion} discusses the implications of our findings and summarizes the main conclusions.

\section{Data} \label{Data}

\subsection{\label{sec:level1}Filament identification}

In order to identify filaments, we have used the 2MASS Redshift Survey (2MRS) \footnote{\url{http://tdc-www.harvard.edu/2mrs/}} galaxies as tracers \citep{Huchra2012}. This catalog contains $43533$ galaxies complete in the K-band up to $K_s=11.75$. The wide sky coverage of the 2MRS ($91 \%$ of the sky) makes it ideal to identify nearby filaments of large angular extent and perform joint studies of the CMB and large-scale structures. \\

We have used galaxies with diameters larger than the median of the distribution ($D>8.5$ Kpc) as tracers, roughly corresponding to galaxies brighter than the $M*$ parameter of the Schechter luminosity function. We have worked on the redshift range $0.004<z<0.040$. The lower limit ($\sim 1200\, km\, s^{-1}$) is adopted in order to exclude galaxies in the very nearby universe where peculiar velocities are relatively large and redshifts do not provide a reliable measure of distance, strongly biasing the identification of filaments. The upper limit corresponds to the threshold where the tracer galaxies drop significantly, also affecting a reliable identification of filaments. \\

We have applied the DisPerSE \citep{Sousbie2011a,Sousbie2011b} code to this sample of galaxies in order to obtain our set of filaments. It is important to clarify that the filaments are defined as the joint of the straight lines (segments) that are traced between two nodes of the filament provided by the algorithm. We also notice that the redshift distribution of 2MRS tracers flattens beyond $z=0.020$ and we will adopt this intermediate threshold to consider two complementary subsamples ($0.004<z<0.020$ and $0.020<z<0.040$). Figure \ref{fig:filament_projection} shows the projected distribution of filaments for both redshift ranges. The persistence parameter given to DisPerSE is $\sigma = 4$; for this value, the fraction of false filaments detected is $\sim 0.006\, \%$ \citep{Sousbie2011a}. The adopted cosmology in this work is $\Omega_{\Lambda} = 0.6911$, $\Omega_M=0.3089$, $\Omega_k = 0$ and $H_0=67.74\, Km\, s^{-1}\, Mpc^{-1}$, consistent with Planck 2018 results \cite{PlanckVI_Cosmological_parameters}.

\subsection{Filament properties}
\label{propiedades}
In order to characterize filaments properly, we have considered two properties: mean linear mass density and the angle between the filament and the line of sight.\\


The former is the ratio between the filament total mass and its length. We estimate a proxy of the filament total stellar mass \cite{Bell2001} by the expected linear relation with the K-band luminosity associated to galaxies within the filaments. For this aim, we sum the K-band luminosity of galaxies within a perpendicular physical distance of $R<4\,$ Mpc (corresponding to $2.71\,$ Mpc/h, where $H_0 = 100\,h Km s^{-1} Mpc^{-1}$, and h=0.6774) from the filament spine. In this way, we obtain the total stellar mass associated to filaments which is roughly proportional to the total filament mass, assuming a constant $M_{tot}$/$M_{\star}$ ratio. Nevertheless, since our aim is to separate the filaments into groups of densest and less dense filaments, these estimates suffices for a suitable division. The length is defined as the sum of the individual lengths of each filament segment.\\

Given the presence of redshift space distortion, whose complete removal is difficult to accomplish, we have used a simple division of filaments into two subsamples, those preferentially along the line of sight, and those preferentially on the plane of the sky. For this aim we define the angle $\phi$ of a filament as the angle that subtends the line joining the extreme points of the filament spine with respect to the line of sight direction to the observer. Thus, $\phi$ is a measure of the global orientation of the filament with respect to the plane of the sky.  This angle may add important information of the effect associated to a given filament since we argue that the interaction between CMB photons and structures accumulates along of the line of sight. Similarly, we also expect a more significant signal for denser filaments, so that the largest contribution to the CMB photons would be associated to the densest and most radially oriented filaments.

\subsection{CMB maps}
We use CMB intensity maps taken by the Planck satellite from the Public Release 3 (PR3) \cite{PlanckI_Overview_and_cosmological_legacy} \footnote{\url{https://pla.esac.esa.int/\#maps}}.  The Planck data were cleaned by 4 different foreground methods as described in Planck Collaboration IV 2018 \cite{PlanckIV_Diffuse_component_separation}. In general, the CMB temperature maps for all $4$ foreground methods agree very well, and in H2025 we made a thorough study of the differences on the CMB cooling in galactic halos, showing only small differences. For this reason, we will here focus on only one of them, namely SMICA (Spectral Matching Independent Component Analysis) for PR3 but use SEVEM (Spectral Estimation Via Expectation Maximisation)  frequency cleaned maps for testing a possible frequency dependence of the signal. We use the Planck common mask created for PR3 \cite{PlanckIV_Diffuse_component_separation} to mask possible galactic residuals as well as point sources, leaving $78\%$ of the sky available for the analysis. To process these maps we use Healpy \cite{Healpix} \footnote{\url{https://healpix.sourceforge.io/}} with a \texttt{Nside} parameter value of 1024. The pixel area for this parameter is $\Omega_p = 11.8026 \ arcmin^2$.

We also use 100 Planck-based simulations, which are publicly available \footnote{\url{https://pla.esac.esa.int/\#maps}} including known Planck systematic effects and noise properties. These simulations will be used to quantify the significance of our results. For a given distance bin, we will define significance as the distance from the CMB data fluctuation temperature to zero CMB fluctuation temperature , given in units of standard deviations.

\section{Methodology}
\label{Methodology}

\subsection{Calculating transversal distances to filaments}

\begin{figure}[htbp]
    \includegraphics[width=0.5\textwidth]{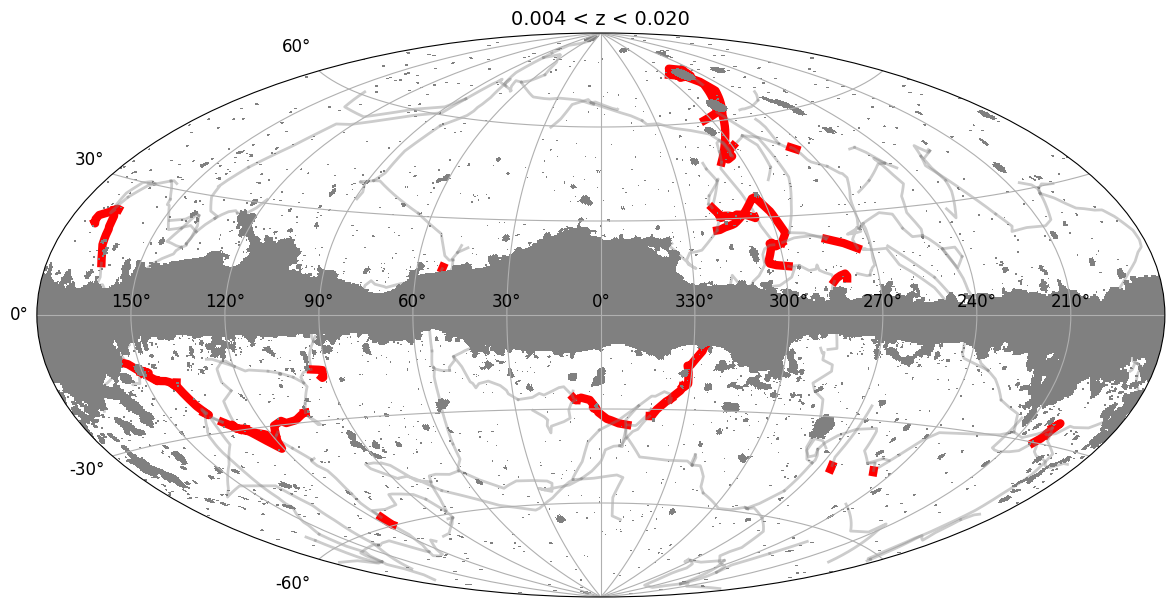}
    \includegraphics[width=0.5\textwidth]{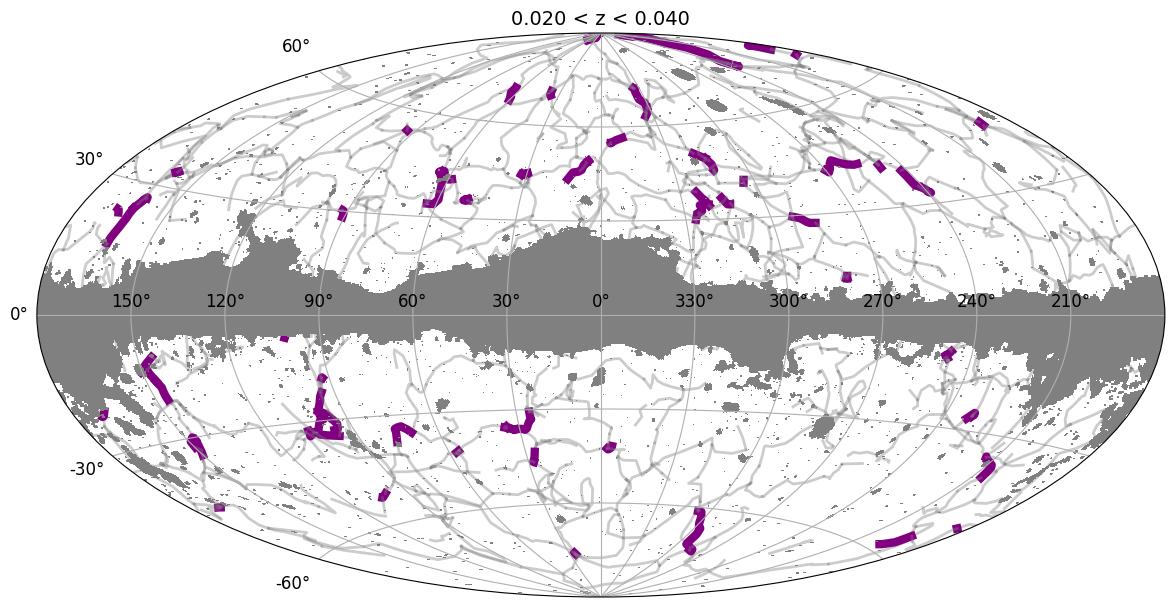}
    \caption{
        Projection of 3D filamentary structure on the plane of the sky in the redshift range $0.004 < z < 0.020$ (upper panel, 278 total filaments identified) and $0.020<z<0.040$ (lower panel, 908 total filaments identified). Grey lines correspond to the sky-projection of all filaments, red lines to the densest hexile and purple to the densest decile in their corresponding redshift interval. The gray mask corresponds to the PR3 CMB common mask, which leaves $\sim 78\% $ of available sky.}
        \label{fig:filament_projection}
\end{figure}

\begin{figure}[htbp]
  \includegraphics[width=0.5\textwidth]{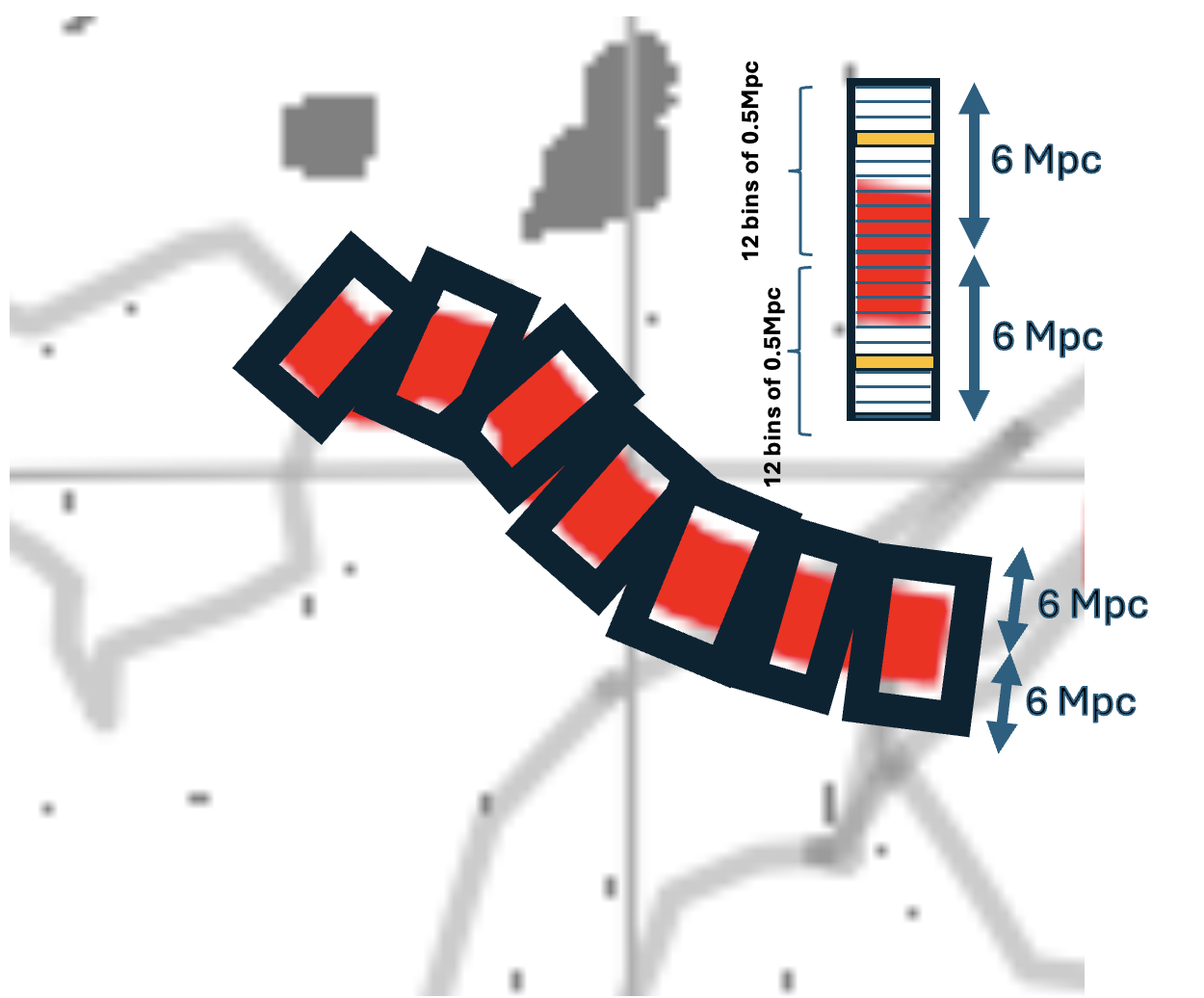}
    \caption{
        The red curved line is a close-up of one of the projected filaments from Figure \ref{fig:filament_projection}. The rectangles along the filament show an example of how the filament is divided into segments, one rectangle for each segment. Each segment has a different length, but the width which is orthogonal to the filament, is always $12$Mpc. In the top of the figure, we show a close-up of the first segment rectangle where we show the 24 distance bins, each of $0.5$Mpc. The mean temperature is taken within each of these bins. The two orange rectangles represent the area $C_{bks}$ of equation \ref{eq:deltat}: the CMB temperature of all the pixels inside the orange rectangles is the contribution of this segment to the given distance bin from the centre of the spine.}
        \label{fig:rectangle}
\end{figure}

Filaments projected on the two-dimensional sphere may produce complex structures as shown in Figure \ref{fig:filament_projection}. In order to calculate distances perpendicular to the filament spine, we developed a method independent of the sky projection structure. To accomplish this, for each projected segment of a filament, we search for all pixels within a rectangle of fixed projected area (see Figure \ref{fig:rectangle}). The length of one of the sides of the rectangle is the length of the filament segment itself and the other is fixed at 6 Mpc at either side of the filament spine as shown in the figure.
This final option is consistent with the scale of the significance of the effect in our previous results and is wider than the typical filament width. For example, Galárraga Espinosa et al. (2020) \cite{GalarragaEspinosa2020} found typical radii of filaments $\sim$ 3–5 Mpc at z=0 using hydrodynamical simulations. We will further divide this rectangle in 12 distance bins of 0.5 Mpc each perpendicular to the filament spine. Since we aim to explore temperature profiles in projected distance, we use the mean segment redshift to transform angular distances into projected Mpc.


\subsection{Multipole removal}
\label{remocion_multipolos}
In H2025, we showed that by removing the largest angular scales of the CMB in simulations (high pass-filtering), the $S/N$ level of the measured mean CMB temperature decrement around galaxies increased significantly. The reason for this is that the angular scale of the galactic filaments is smaller than the largest CMB fluctuations. The CMB flucuations at large scale act as noise to the smaller scale fluctuations associated to the galaxies which we consider the signal. We found in H2025 that, for galaxies $z<0.02$, by removing the 5 first multipole moments of the CMB maps, we could optimize $S/N$ meaning that we increase the amplitude of the small scale fluctuations including a possible signal from galactic filaments with respect to the amplitude of the large scale CMB background. Here we will also consider filaments in the redshift range $0.02<z<0.04$. A typical massive filament has a total length of about $15-20$Mpc which corresponds to an angular extention of about $6-9$ degrees in the sky at these redshifts, but as most of these are curved (as we can see in Fig. \ref{fig:filament_projection}), their actual angular extension is typically about 5 degrees corresponding to multipoles $\ell\gtrsim32$. We therefore choose to remove all multipoles $\ell\leq32$ in both Planck data and simulations for analysis in the redshift range $0.02<z<0.04$. For the larger filaments in $z<0.02$ we will remove $\ell\leq5$ as in H2025. To test consistency, we will also apply the analysis to maps with $ \ell>5, \, \ell>10, \, \ell>16, \, \ell>32, \, \ell>64$ and $\ell>128$.

The method for estimating the lowest multipole moments outside the mask, and then subtracting is explained in detail in H2025. In short, for the lowest multipole removals $l\leq16$, we calculate and invert the full $a_{\ell m}$ coupling matrix with the given mask and apply this to the cut-sky $a_{\ell m}$. This is the same method used for mono and dipole removal in the Healpix package \citep{Healpix}. For removing higher multipoles, we use a simplified procedure where we downgrade and smooth the map to the resolution corresponding to the maximum multipole to remove. In H2025 we found that the two methods give similar results.

\subsection{Transversal temperature profiles}
The previously detected CMB temperature decrement reported in L2023 and H2025 is an order of magnitude smaller than the standard deviation of the CMB temperature fluctuations. We therefore stack CMB temperature profiles transverse to the filaments in order to reach a sufficiently high $S/N$ level. In practice, this stacked profile is obtained by averaging the transverse CMB temperature profiles of each individual filament.
We recall that CMB pixels in the CMB common mask are not being used in our statistical analysis.

The mean CMB temperature $\langle \Delta T\rangle(d_\perp)$ at a perpendicular distance $d_\perp$ from the filamentary spine taken over all the filaments is calculated by
\begin{equation}
    \Delta T(d_\perp^b) = \frac{1}{N_{fil}}
    \sum_{k=1}^{Nfil}\left(\frac{1}{N_{bk}} \sum_{s} \sum_{i \in C_{bks}} \Delta T(k,s,i)\right)
    \label{eq:deltat}
\end{equation}

We notice that when using this equation, we are already constructing a CMB temperature profile with 12 equidistant bins on each side perpendicular to a segment in the filament spine up to 6 Mpc (see Figure \ref{fig:rectangle}). The equation reads as: the k-index is for filaments, the s-index is for the segments of a given filament, and the i-index is for pixels within distance bin $b$, so we construct the CMB temperature profile by first adding up all the CMB temperatures associated to pixels inside the corresponding distance bin \textit{b} (i.e. all pixels inside the area $C_{bks}$ represented by the two orange rectangles in figure \ref{fig:rectangle}) and divide by the total number of pixels that fall inside bin b, across all segments of filament k ($N_{bk}$). Then we repeat this process for each segment, and finally for each filament, dividing at the end by the number of filaments.
The reason for firstly taking mean temperature over each filament is
to avoid the filaments with many pixels in a given distance bin to dominate the signal. It is important to clarify that, in our method, each pixel appears at most once
per filament. However, since two or more filaments may
overlap in their projection onto the sky, pixels that are
shared will be considered independently for each filament
and could contribute to the profile at different perpendicular distances ($d_{\perp} < 6$ Mpc). \\
We have used a bootstrap resampling technique method to asses the intrinsic uncertainty of the temperature profiles obtained. This was accomplished by applying $500$ iterations of random resampling to the temperature profiles calculated, in order to obtain the standard deviation of the derived profiles.

\section{Results}
\label{Results}

In this section we first present the results in the redshift range $0.004<z<0.020$ which was also used in L2023 and H2025. Then, we extend our analysis to the range $0.020<z<0.040$ testing whether the cooling effect seen in the first redshift range may be visible also at these higher redshifts. Finally, we also consider the total redshift interval $0.004<z<0.040$. As detailed above, we will present our main results based on maps with $\ell>5$ for the range $0.004<z<0.020$, and $\ell>32$ in the more distant ranges $0.020<z<0.040$. For the full range $0.004<z<0.040$ results of the variance weighted combination of the results of the individial ranges will be presented along with a combined analysis of all filaments in the full range applied to the same CMB map with different multipole cuts.

\subsection {Subsample percentiles according to linear luminosity density}
\begin{figure}[htbp]
    \includegraphics[width=0.5\textwidth]{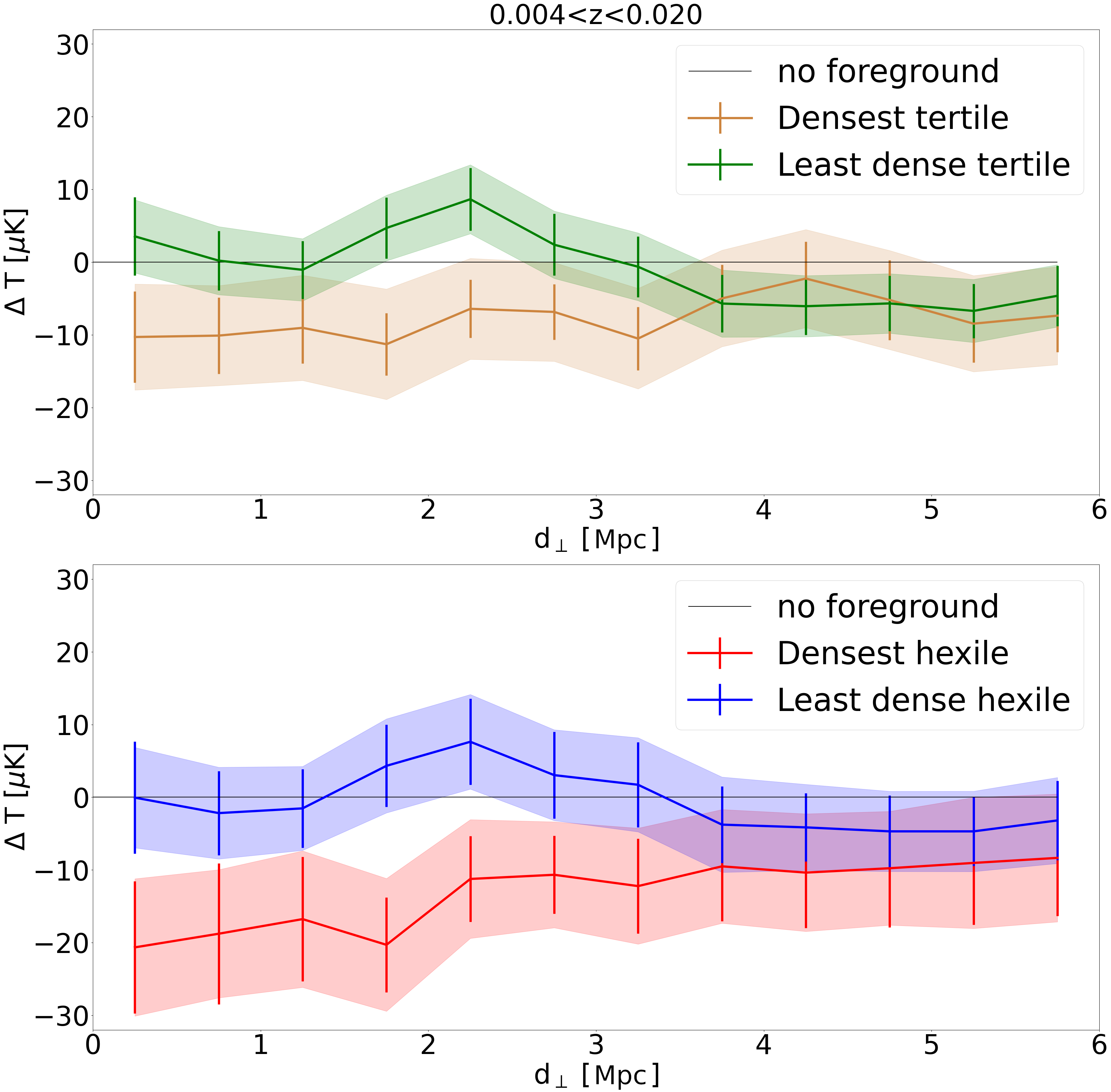}
    \caption{Mean CMB temperature profiles in perpendicular distances to the filament spine for the redshift range $0.004<z<0.020$, with $\ell>5$. The solid line is the data, vertical solid lines are the standard deviation of the data and the shadowed regions are the variance in synthetic simulations. In the upper panel the analysis is done for the filaments in the densest and least dense tertile of luminosity density. In the lower panel the analysis is done for the filaments in the densest and least dense hexile of luminosity density.}
        \label{fig:fig2}
\end{figure}

Firstly, we divide the filaments in tertiles of linear luminosity density and estimate CMB temperature profiles for the densest as well as the least dense tertile. This is shown in the upper panel of Figure \ref{fig:fig2}.

\begin{figure}[htbp]
    \includegraphics[width=0.45\textwidth]{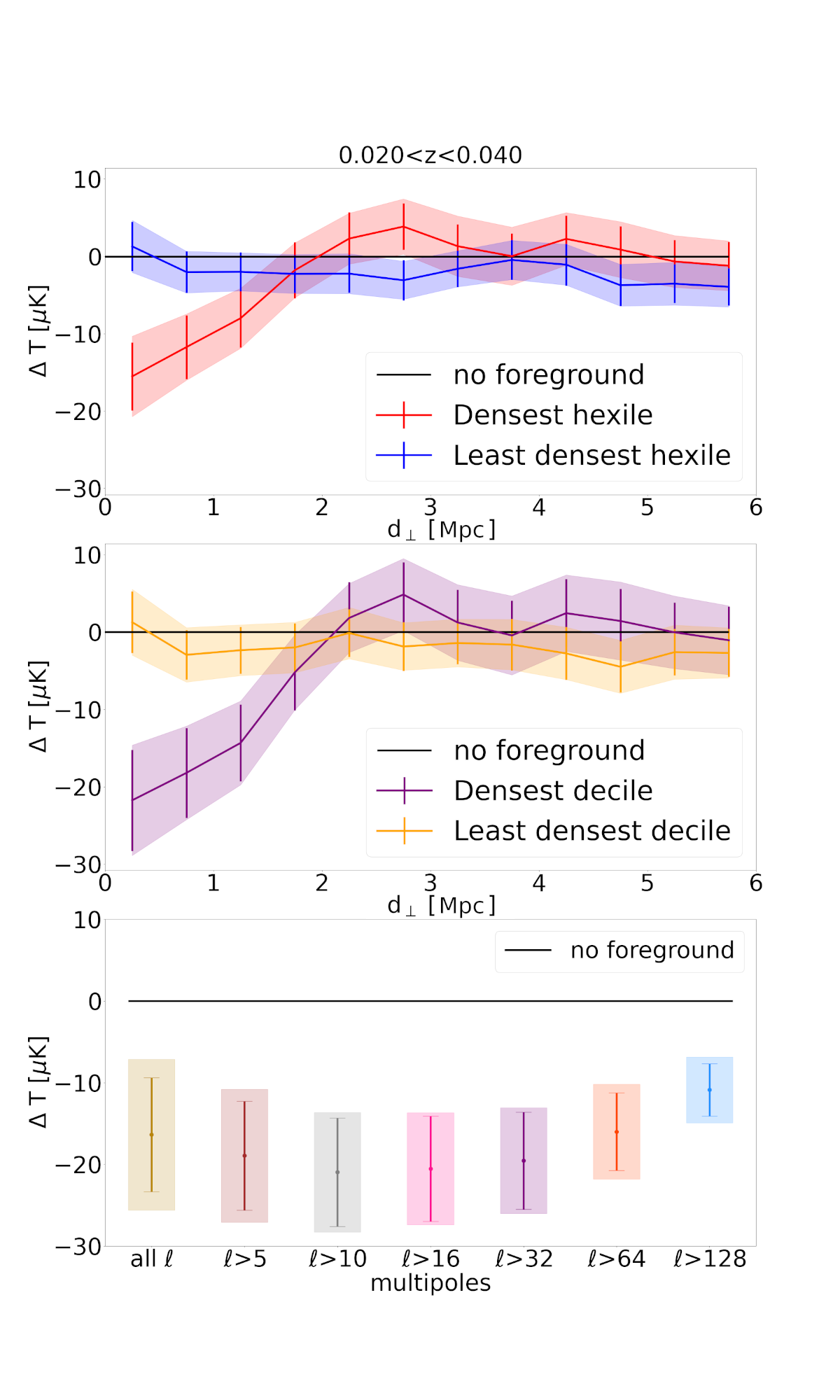}
    \caption{Analysis of filaments in the redshift range $0.020<z<0.040$. Upper panel: densest and least dense (luminosity per unit length) hexile filaments. Middle panel: densest and least dense decile filaments. In both panels, the solid line is the data, vertical solid lines are the standard deviation of the data and the shadowed regions are the variance in CMB synthetic simulations. Lower panel: mean CMB temperature of a bin ($0-1.0\,$Mpc) of densest filaments decile, for different ranges of low multipoles removed. The solid vertical lines correspond to the standard deviation of the data, while the shadowed boxes correspond to the variance from CMB synthetic simulations with the same ranges of multipoles removed.}
        \label{fig:fig5}
\end{figure}

We can see that the values of the CMB temperature in pixels associated to the least dense tertile of filaments are consistent with the average CMB values. On the other hand, those associated to the densest tertile show a CMB temperature decrement, presenting a systematic trend with the perpendicular distance to the filament spine. One could argue that this effect is mainly associated to the galaxies (as shown in L2023 and H2025). With the aim of addressing the contribution of individual galaxies in the filaments, we have removed the pixels within $150$ projected Kpc from all galaxies in the 2MRS catalog in the redshift range of interest. No significant difference in the transversal CMB temperature profiles was seen. We therefore conclude that the foreground effect is not restricted only to galaxies, but is also associated to the large-scale structures traced by the filaments. \\

In the lower panel we show the results for filaments belonging to the highest and lowest luminosity density hexile. We can see again a CMB temperature decrement signal for the densest hexile (red line in lower panel) with higher significance respect to the densest tertile (light brown in upper panel). As expected, the significance increases with increasing luminosity density, but we are still far from the significances in H2025. In H2025, the dependency of galaxy morphology and in particular on galaxy mass was taken into account as the CMB temperature was only measured in the late type spiral galaxies with a mass larger than the median mass. In our present approach, although we focus on the more massive filaments, we use all parts of the filaments, independent of the local variations of mass density along the filament. We argue that the less massive parts of the filaments dilute the signal and give rise to a less significant detection when using the filament centered profile instead of the galaxy centered profile. Also, in H2025, by focusing on massive galaxies and thereby including very dense regions of less massive filaments as well as large overdensities outside the filaments, we also obtained more statistics. The filament based approach is however better for larger distances which is our main focus here, as properties of individual galaxies are less certain at higher redshifts.\\

As a next step, we have considered a significantly larger volume corresponding to the redshift range $0.020<z<0.040$. In this redshift range, there is a total of $908$ filaments compared to the previous redshift range where a total number of $278$ filaments were found. This larger sample of filaments allows for a larger number of sub-samples corresponding to different percentiles. The results are shown in the upper (middle) panels of Figure \ref{fig:fig5} for the densest and least densest hexile (decile). We can see for this extended redshift range that the signal is restricted to the first bins in transversal distances. We argue that this behavior is related to the smaller angular scale subtended by the filament width at these higher redshifts. As in the previous range, we have also tested the removal of pixels within $150$ projected Kpc from the 2MRS galaxies and again there were only tiny differences in the profiles. \\

The lower panel in this figure shows the mean CMB temperature within $1$Mpc as a function of minimum $\ell$ of the CMB map. We can see that for larger multipole removal, the variance in the synthetic simulation are similar to the data bootstrap-derived standard deviation. For lower values of minimum $\ell$, the signal and variance is affected by large angular scales CMB temperature fluctuation. For larger multipole cuts, the CMB temperature in the filaments is gradually getting less negative, but as the variance of the CMB flucuations are decreasing with increasing multipole cut, the relative temperature difference with respect to the CMB and therefore also the significance of the signal, is almost unchanged.\\

In the upper part of table \ref{tab:significancias_decil_lejano} the significances for the densest decile in this redshift interval, up to $2$ Mpc, for different multipole removals is shown. We see that for the first distance bin within $<0.5$Mpc, we reach $\sim3\sigma$ significance. At a distance of $<1.5$Mpc there is still a significant $2.5\sigma$ signal, whereas the signal diappers quite abruptly after this as was also seen in the figures. While the maximum significance is found as expected for this redshift range at $\ell>32$, the signal is very consistent also for other multipole cuts.\\

\begin{figure}[htbp]
    \includegraphics[width=0.45\textwidth]{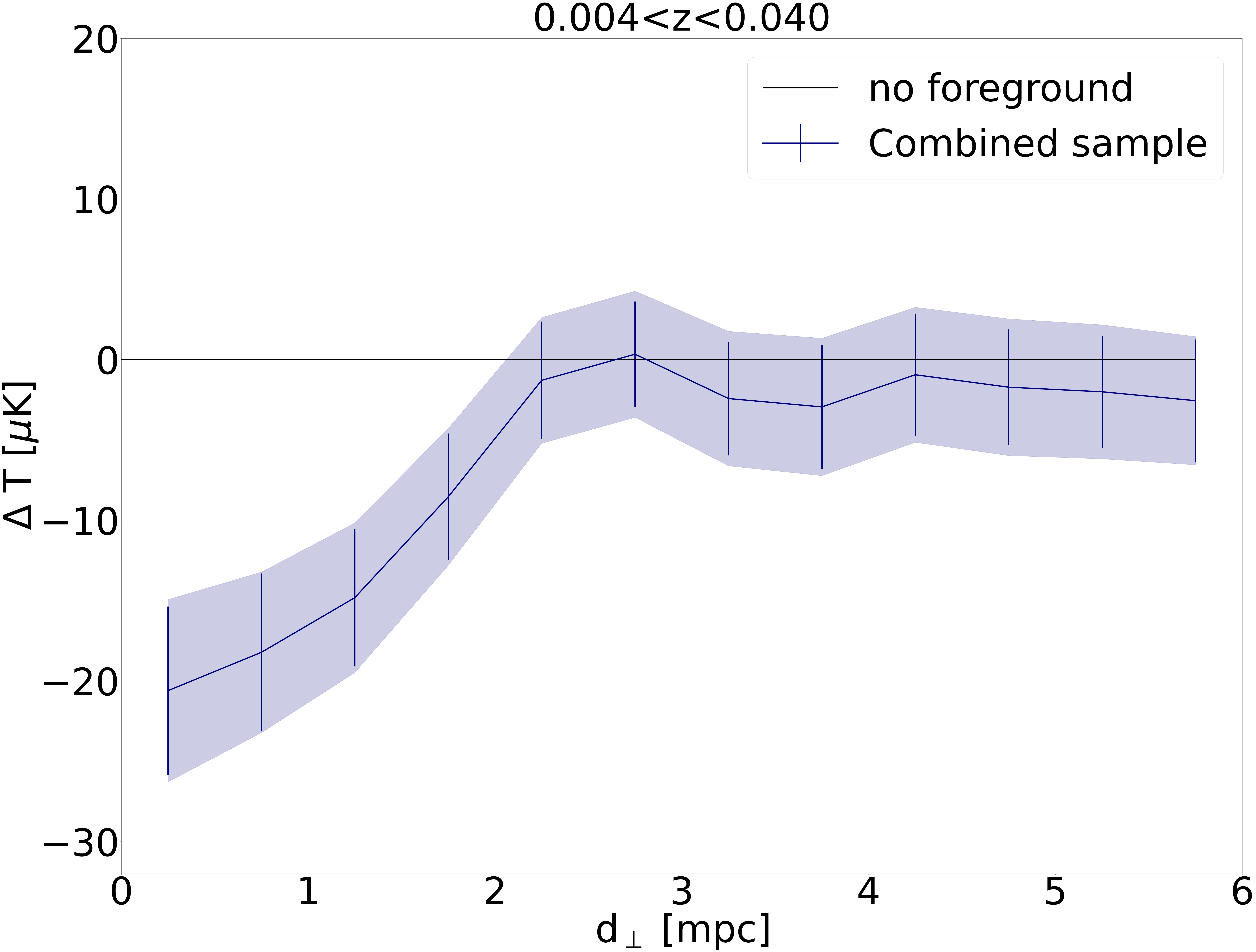}
    \caption{CMB temperature profile for the full redshift range $0.004<z<0.04$ as variance weighted combination of the individual ranges based on the densest hexile with $\ell>5$ for the lowest redshift range and densest decile with $\ell>32$ for the highest redshift range. The solid line is the data, the vertical solid line is the standard deviation of this data and the shadowed regions are the variance in simulations.}
        \label{fig:quintilYdecil_sinang_rango_completo}
\end{figure}

Finally, we explore the mean CMB temperature taken over all filaments in the complete redshift range $0.004<z<0.04$. In Figure \ref{fig:quintilYdecil_sinang_rango_completo} we show the CMB temperature profile for the variance weighted average between the results for the optimal case in each redshift range, the densest hexile for $0.004<z<0.02$ using the multipole cut $\ell>5$ and the densest decile for $0.02<z<0.04$ removing $\ell\leq32$. We find a $\sim3.5\sigma$ detection for the full range.

\begin{table}[]
\begin{tabular}{|l|lllllll|}
\hline
 &
  \multicolumn{7}{l|}{Significance ($\sigma$), all angles} \\ \hline
$d_{\perp}$ {[}Mpc{]} &
  \multicolumn{1}{l|}{all $\ell$} &
  \multicolumn{1}{l|}{$\ell$\textgreater{}5} &
  \multicolumn{1}{l|}{$\ell$\textgreater{}10} &
  \multicolumn{1}{l|}{$\ell$\textgreater{}16} &
  \multicolumn{1}{l|}{$\ell$\textgreater{}32} &
  \multicolumn{1}{l|}{$\ell$\textgreater{}64} &
  $\ell$\textgreater{}128 \\ \hline
0.0 - 0.5 &
  \multicolumn{1}{l|}{1.79} &
  \multicolumn{1}{l|}{2.30} &
  \multicolumn{1}{l|}{2.77} &
  \multicolumn{1}{l|}{2.88} &
  \multicolumn{1}{l|}{2.90} &
  \multicolumn{1}{l|}{2.64} &
  2.49 \\ \hline
0.5 - 1.0 &
  \multicolumn{1}{l|}{1.63} &
  \multicolumn{1}{l|}{2.22} &
  \multicolumn{1}{l|}{2.81} &
  \multicolumn{1}{l|}{2.95} &
  \multicolumn{1}{l|}{2.96} &
  \multicolumn{1}{l|}{2.68} &
  2.46 \\ \hline
1.0 - 1.5 &
  \multicolumn{1}{l|}{1.24} &
  \multicolumn{1}{l|}{1.92} &
  \multicolumn{1}{l|}{2.50} &
  \multicolumn{1}{l|}{2.57} &
  \multicolumn{1}{l|}{2.64} &
  \multicolumn{1}{l|}{2.22} &
  1.58 \\ \hline
1.5 - 2.0 &
  \multicolumn{1}{l|}{0.20} &
  \multicolumn{1}{l|}{0.70} &
  \multicolumn{1}{l|}{1.12} &
  \multicolumn{1}{l|}{1.04} &
  \multicolumn{1}{l|}{1.09} &
  \multicolumn{1}{l|}{0.45} &
  1.03 \\ \hline
  &
  \multicolumn{7}{l|}{Significance ($\sigma$), radially oriented} \\ \hline
  0.0 - 0.5 &
 \multicolumn{1}{l|}{2.33} &
  \multicolumn{1}{l|}{2.56} &
  \multicolumn{1}{l|}{3.12} &
  \multicolumn{1}{l|}{3.19} &
  \multicolumn{1}{l|}{3.22} &
  \multicolumn{1}{l|}{2.98} &
  2.89 \\ \hline
  0.5 - 1.0 &
\multicolumn{1}{l|}{2.26} &
  \multicolumn{1}{l|}{2.55} &
  \multicolumn{1}{l|}{3.30} &
  \multicolumn{1}{l|}{3.39} &
  \multicolumn{1}{l|}{3.47} &
  \multicolumn{1}{l|}{3.26} &
  3.01 \\ \hline
  1.0 - 1.5 &
\multicolumn{1}{l|}{1.59} &
  \multicolumn{1}{l|}{1.90} &
  \multicolumn{1}{l|}{2.58} &
  \multicolumn{1}{l|}{2.67} &
  \multicolumn{1}{l|}{2.80} &
  \multicolumn{1}{l|}{2.44} &
  1.46 \\ \hline
1.5 - 2.0 &
  \multicolumn{1}{l|}{0.41} &
  \multicolumn{1}{l|}{0.62} &
  \multicolumn{1}{l|}{1.13} &
  \multicolumn{1}{l|}{1.15} &
  \multicolumn{1}{l|}{1.20} &
  \multicolumn{1}{l|}{0.71} &
  0.97 \\ \hline
\end{tabular}

\caption{Significances for the filaments of the densest decile of luminosity density in the redshift range $0.020<z<0.040$ per $0.5$ Mpc bin. We consider the cases: all $\ell$, $\ell>5$, $\ell>10$, $\ell>16$, $\ell>32$, $\ell>64$ and $\ell>128$. Upper part of table: including all filaments in the decile (as in figure \ref{fig:fig5}, middle panel). Lower part: including only radially oriented filaments (as in figure \ref{fig:sextilYdecil_angulos}, lower panel)  }
\label{tab:significancias_decil_lejano}
\end{table}

\subsection{Orientation of filaments with respect to the line of sight}
In the previous analysis, we have shown that the densest filaments have the strongest effect on the CMB photons. Thus, we expect that those filaments oriented along the direction of the line of sight will also exhibit a stronger effect since, in this case, photons travel larger distances in filament environments, which could cause a deeper temperature profile in contrast to those preferentially on the plane of the sky. \\

Here we consider the densest sub-samples of all the previous redshift ranges and analyze these filaments taking into account the angle $\phi$ defined in section \ref{propiedades} between the filament and the line of sight. We divide the sample into radial filaments ($\phi < 60^{^\circ}$) and plane-of-the-sky filaments ($\phi > 60^{^\circ}$). We acknowledge the fact that the tracer galaxies from which the filaments are identified are subject to redshift-space distortions. Therefore, the angle $\phi$ does not exactly reflect the true angle between the filament and the line of sight direction. However, we argue that it is nevertheless a suitable parameter for a simple division of the sample of filaments into these categories. We find that the decrement signal detected is not extremely sensitive to a particular threshold angle dividing into radial and non radial filaments. 

\begin{figure}[htbp]
    \includegraphics[width=0.4\textwidth]{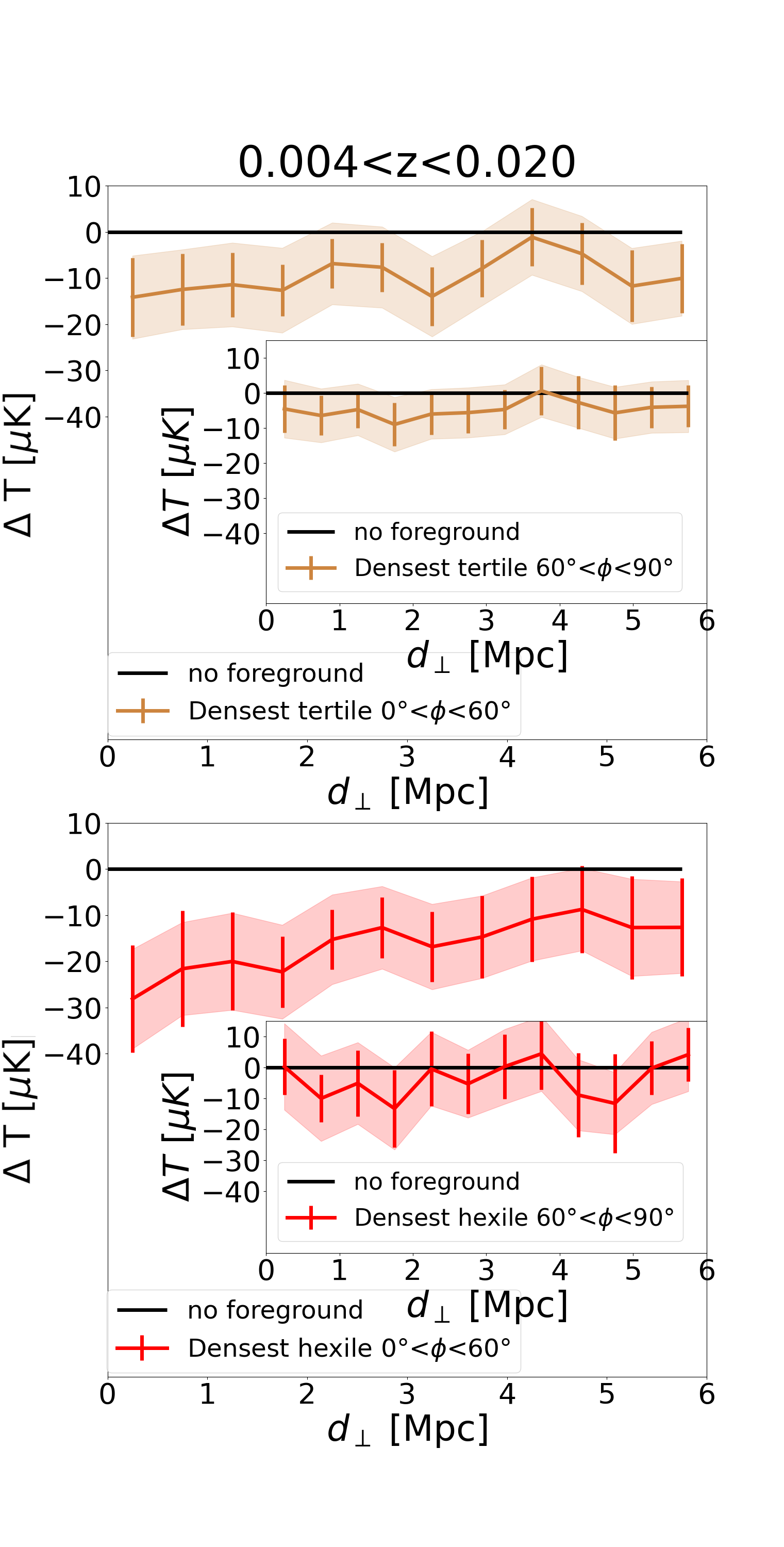}
    \caption{Filaments in the redshift range $0.004<z<0.020$, with $\ell>5$ for the densest tertile (upper panel) and densest hexile (lower panel) for the preferentially radially oriented filaments ($\phi < \ang{60}$). 
    The inset plot corresponds to the filaments oriented preferentially on the plane of the sky ($\phi > \ang{60}$). The solid line is the data, the vertical solid line is the standard deviation of this data and the shadowed regions are the variance in simulations.}
        \label{fig:tercilYsextil_angulo}
\end{figure}
\begin{figure}[htbp]
    \includegraphics[width=0.4\textwidth]{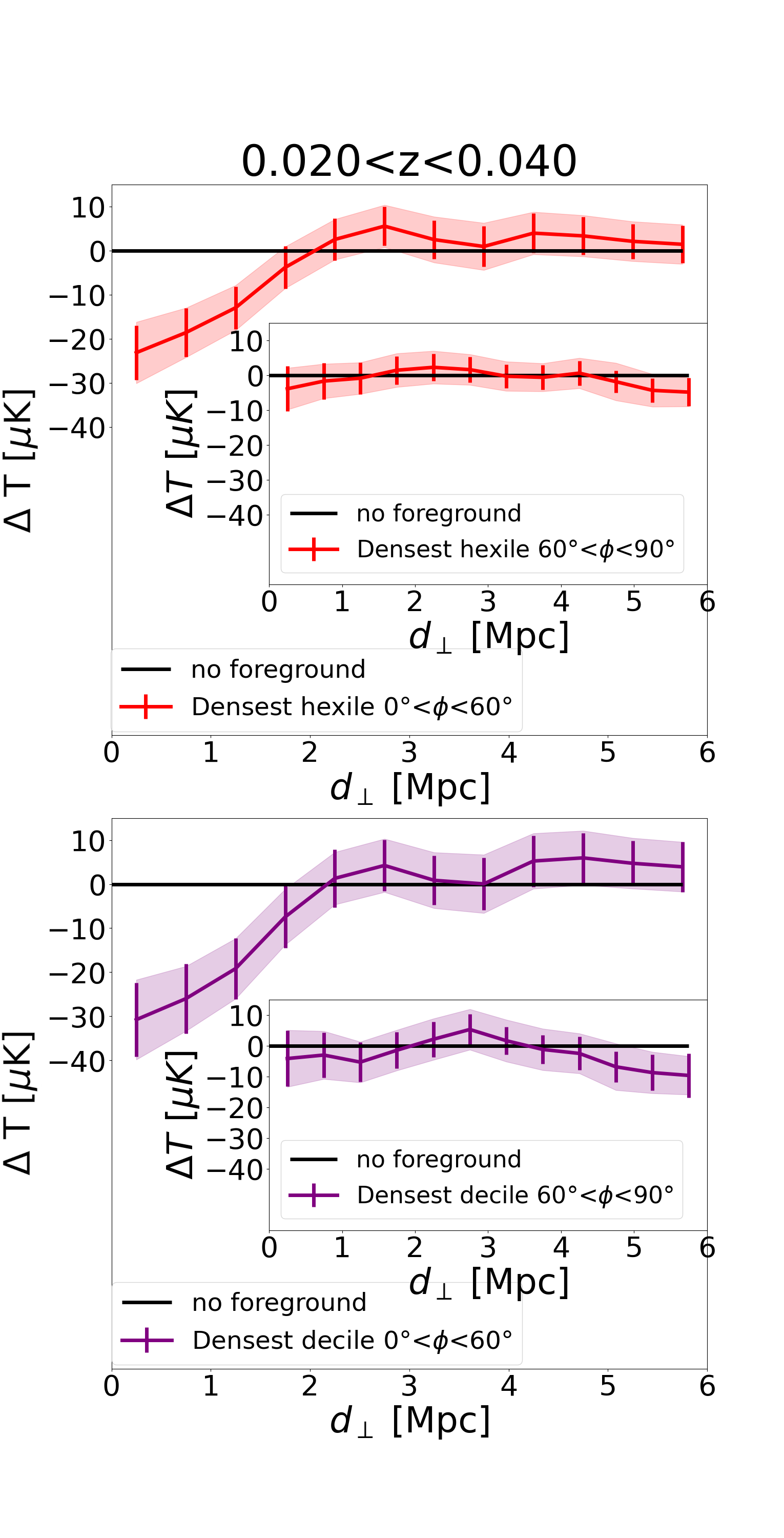}
    \caption{Filaments in the redshift range $0.020<z<0.040$, with $\ell>32$ for the densest hexile (upper panel) and densest decile (lower panel) for the preferentially radially oriented filaments ($\phi < \ang{60}$). 
    The inset plot corresponds to the filaments oriented preferentially on the plane of the sky ($\phi > \ang{60}$). The solid line is the data, the vertical solid line is the standard deviation of this data and the shadowed regions are the variance in simulations.}
        \label{fig:sextilYdecil_angulos}
\end{figure}
\begin{figure}[htbp]
    \centering
    \includegraphics[width=0.48\textwidth]{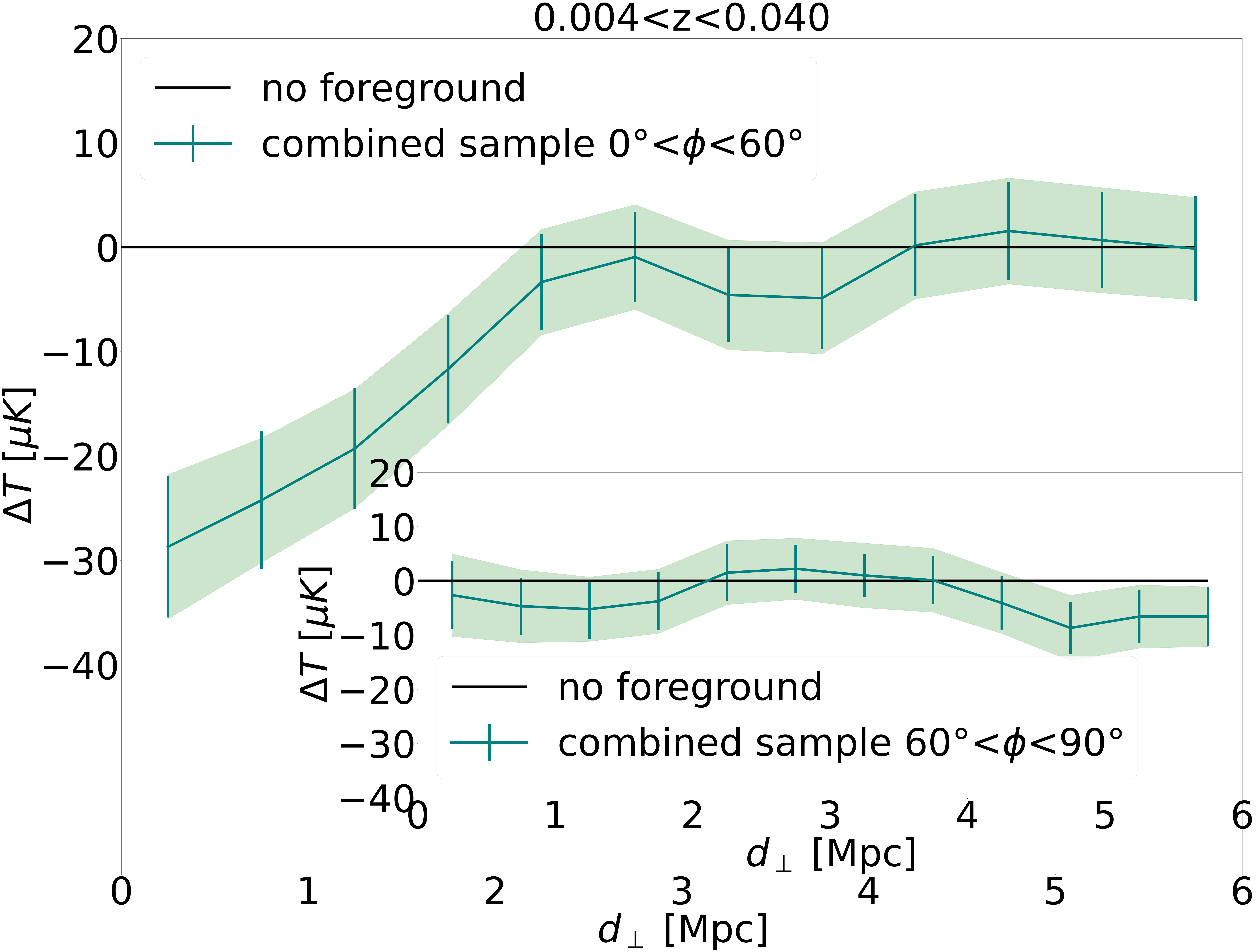}
    \caption{CMB temperature profile for the full redshift range $0.004<z<0.04$ for the preferentially radially oriented filaments ($\phi < \ang{60}$). The results for the full range is obtained as a variance weighted combination of the individual ranges based on the densest hexile with $\ell>5$ for the lowest redshift range and densest decile with $\ell>32$ for the highest redshift range.  The inset plot corresponds to the filaments oriented preferentially on the plane of the sky ($\phi > \ang{60}$). The solid line is the data, the vertical solid line is the standard deviation of this data and the shadowed regions are the variance in simulations.}
    \label{fig:quintilYdecil_conangulo}
\end{figure}

In figure \ref{fig:tercilYsextil_angulo}, we show results for the closest redshift interval ($0.004<z<0.020$) and $\ell > 5$ and figure \ref{fig:sextilYdecil_angulos} for the more distant redshift range ($0.020<z<0.040$) and $\ell > 32$. Lastly, figure \ref{fig:quintilYdecil_conangulo} correspond to the result for the complete redshift interval ($0.004<\ell<0.040$. It is remarkable how the radially oriented filaments in all cases (main panels) show a strong (up to $4\sigma$) signal, while there is a corresponding lack of signal for those filaments in the plane of the sky (inset panels). We argue that this is a clear indication that the extragalactic foreground is associated to 3-D large-scale structures along the radial direction, consistent with the arguments given at the beginning of this section.\\

Finally, to asses the robustness of the results, in the lower part of Table \ref{tab:significancias_decil_lejano} as well as in Table \ref{tab:significancias} we show the significance for the densest decile of the decrement signals within $2$ Mpc from the filaments for radially oriented filaments. The first table show results for the distant redshift range, while the latter shows the full $0.004<z<0.040$ range, for various multipole cuts. Note that for the full range, we did not analyze the CMB temperature profiles for all filaments in the full range on the same map, but rather combine results from the individual redshift ranges.
The last column shows the corresponding result for the variance weighted combination. We see that the radially oriented filaments show $3-4\sigma$ significances in the first bin. Not unexepectedly, the variance weighted combination of the redshift ranges shows the highest signficance at $4.1\sigma$, as in this case the multipole range optimizing $S/N$ for the angular size of filaments in each redshift range is taken into account.



\begin{table}[]
\begin{tabular}{|l|llllllll|}
\hline
      & \multicolumn{8}{l|}{Significance ($\sigma$)}                                                        \\ \hline
$d_{\perp}$ {[}Mpc{]} & \multicolumn{1}{l|}{all $\ell$} & \multicolumn{1}{l|}{$\ell$\textgreater{}5} & \multicolumn{1}{l|}{$\ell$\textgreater{}10} & \multicolumn{1}{l|}{$\ell$\textgreater{}16} & \multicolumn{1}{l|}{$\ell$\textgreater{}32} & \multicolumn{1}{l|}{$\ell$\textgreater{}64} & \multicolumn{1}{l|}{$\ell$\textgreater{}128} & \multicolumn{1}{l|}{comb} \\ \hline
0.0 - 0.5 & \multicolumn{1}{l|}{2.74} & \multicolumn{1}{l|}{3.18} & \multicolumn{1}{l|}{3.34} & \multicolumn{1}{l|}{3.38} & \multicolumn{1}{l|}{3.51} & \multicolumn{1}{l|}{3.29} &  \multicolumn{1}{l|}{3.45} & 4.14 \\ \hline
0.5 - 1.0 & \multicolumn{1}{l|}{2.65} & \multicolumn{1}{l|}{3.20} & \multicolumn{1}{l|}{3.52} & \multicolumn{1}{l|}{3.56} & \multicolumn{1}{l|}{3.75} & \multicolumn{1}{l|}{3.63} &  \multicolumn{1}{l|}{3.46} & 4.07 \\ \hline
1.0 - 1.5 & \multicolumn{1}{l|}{2.19} & \multicolumn{1}{l|}{2.72} & \multicolumn{1}{l|}{3.04} & \multicolumn{1}{l|}{3.10} & \multicolumn{1}{l|}{3.25} & \multicolumn{1}{l|}{2.98} &  \multicolumn{1}{l|}{2.18} & 3.39 \\ \hline
1.5 - 2.0 & \multicolumn{1}{l|}{1.46} & \multicolumn{1}{l|}{1.77} & \multicolumn{1}{l|}{1.97} & \multicolumn{1}{l|}{2.10} & \multicolumn{1}{l|}{2.15} & \multicolumn{1}{l|}{1.55} &  \multicolumn{1}{l|}{0.1} & 2.17 \\ \hline
\end{tabular}
\caption{Significances for the filaments of the densest decile, radially oriented, in the total redshift range $0.004<z<0.040$ (figure \ref{fig:quintilYdecil_conangulo}) per $0.5$ Mpc bin. We consider the cases: all $\ell$, $\ell>5$, $\ell>10$, $\ell>16$, $\ell>32$, $\ell>64$ and $\ell>128$. The last column corresponds to the variance weighted combination of the individual ranges based on the densest hexile with $\ell>5$ for the lowest redshift range and densest decile with $\ell>32$ for the highest redshift range.}
\label{tab:significancias}
\end{table}


\section{Discussion}
\label{Discussion}
Cosmic filaments are common anisotropic structures in the universe, which have been largely studied in both observations and cosmological simulations. The recently detected extragalactic CMB foreground is an unexplained CMB temperature decrement effect associated with galaxies extending to several times the galaxy halo scales (H2023). The geometrically large extension and narrow width of the filaments provide a natural cylindrical coordinate system along the filament spine for studying the possible dependence of the foreground effect from filaments. We have therefore analyzed the mean CMB temperature decrement of the SMICA map pixel temperatures as a function of the cylindrical radial coordinate perpendicular to the filament spine in samples obtained in the nearby universe as traced by the 2MRS galaxies.\\

In this paper, by using CMB temperature profiles around nearby galactic filaments instead of the galaxies themselves, we have discovered several new properties of this unknown CMB cooling mechanism and confirmed its existence independently through the analysis of new and independent galaxy samples. We can summarize these new findings in four main points:
\begin{itemize}
\item Previous papers have detected the CMB temperature decrement mainly in and around the galactic halos and attributed extended signals to neighbouring galaxies (but see \cite{Cruz} where the signal is correlated to the dark matter density). Here we show that there seem to be a CMB decrement signal associated to the filamentary structure itself and thereby possibly to the density field. By masking a radius of $150$Kpc around the galaxies, the filamentary CMB temperature decrement profiles are almost unchanged.
\item Previous detections of the cooling of CMB photons in galaxies have been limited to the very nearest universe $z<0.02$. Here we analyze the next redshift shell $0.02<z<0.04$ and find, using the CMB temperature profile around filaments, an even stronger signal than in $z<0.02$. The effect is now confirmed independently in a new redshift range at the $3-4\sigma$ level.
\item While some dependence of the effect on galactic mass has been discussed in previous works, here we show a very strong dependence on the filamentary density. We have considered a measure of the K-band luminosity density per unit length within 4 Mpc from the filament spine as a proxy to the filament mass density per unit length. Such measure provides a parameter to distinguish between the densest filaments which trace the large scale structure, and the less dense which more generally reside in void environments. Thus, both the mass content of these two filament types and their dynamical stage are expected to differ significantly. We find a clear trend of larger temperature decrements the higher the filamentary mass and conversely, no signal in the less dense filaments.
\item We have taken into account the 3-D shape of filaments and have explored the fact that it would be expected that the filaments produce a larger signal when they are oriented along the line of sight with respect to those on the plane of the sky. The results of these tests regarding the orientation of filaments are also conclusive, most of the signal comes from filaments preferentially oriented along the line of sight, giving an additional strong argument for the association of the foreground detection to filaments.
\end{itemize}

\begin{figure}[htbp]
  \includegraphics[width=0.4\textwidth]{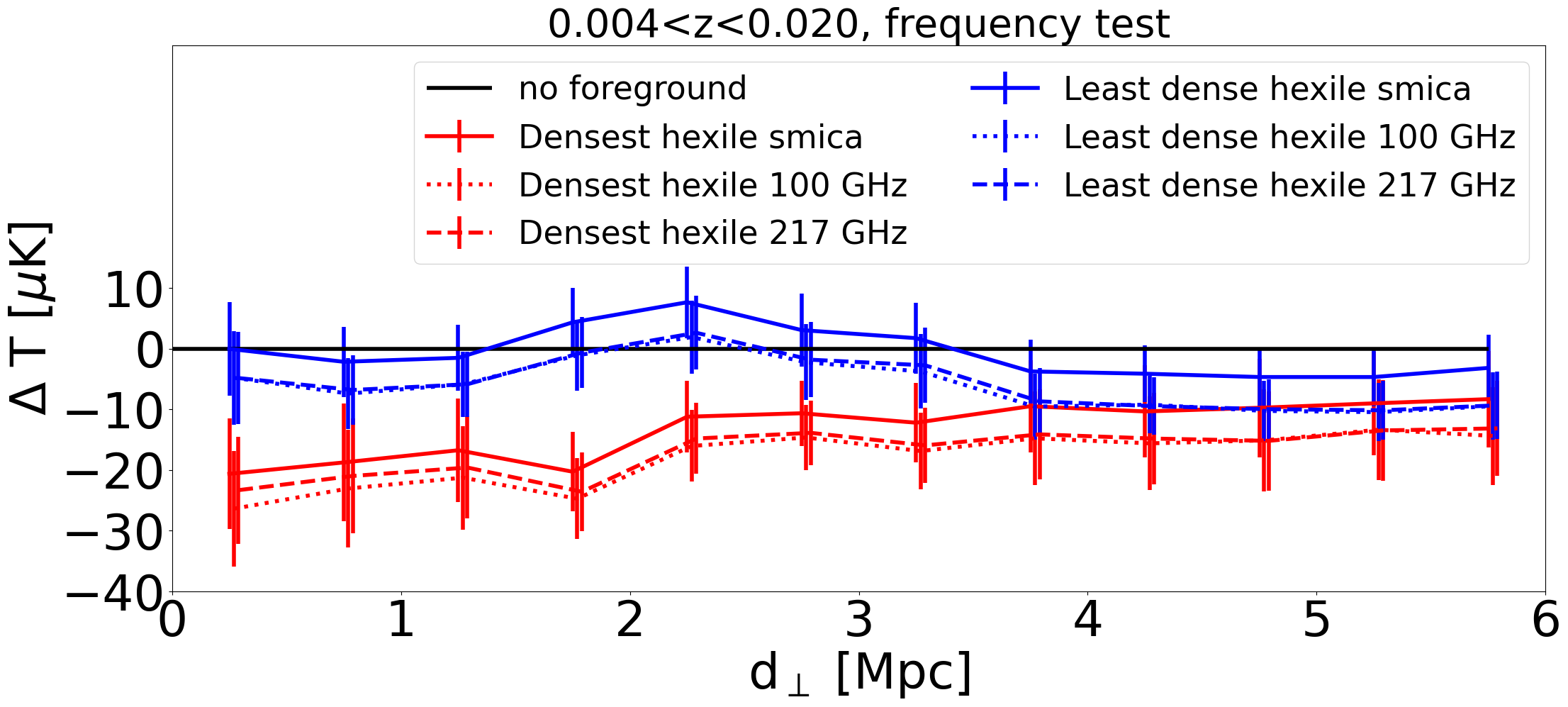}
  \includegraphics[width=0.4\textwidth]{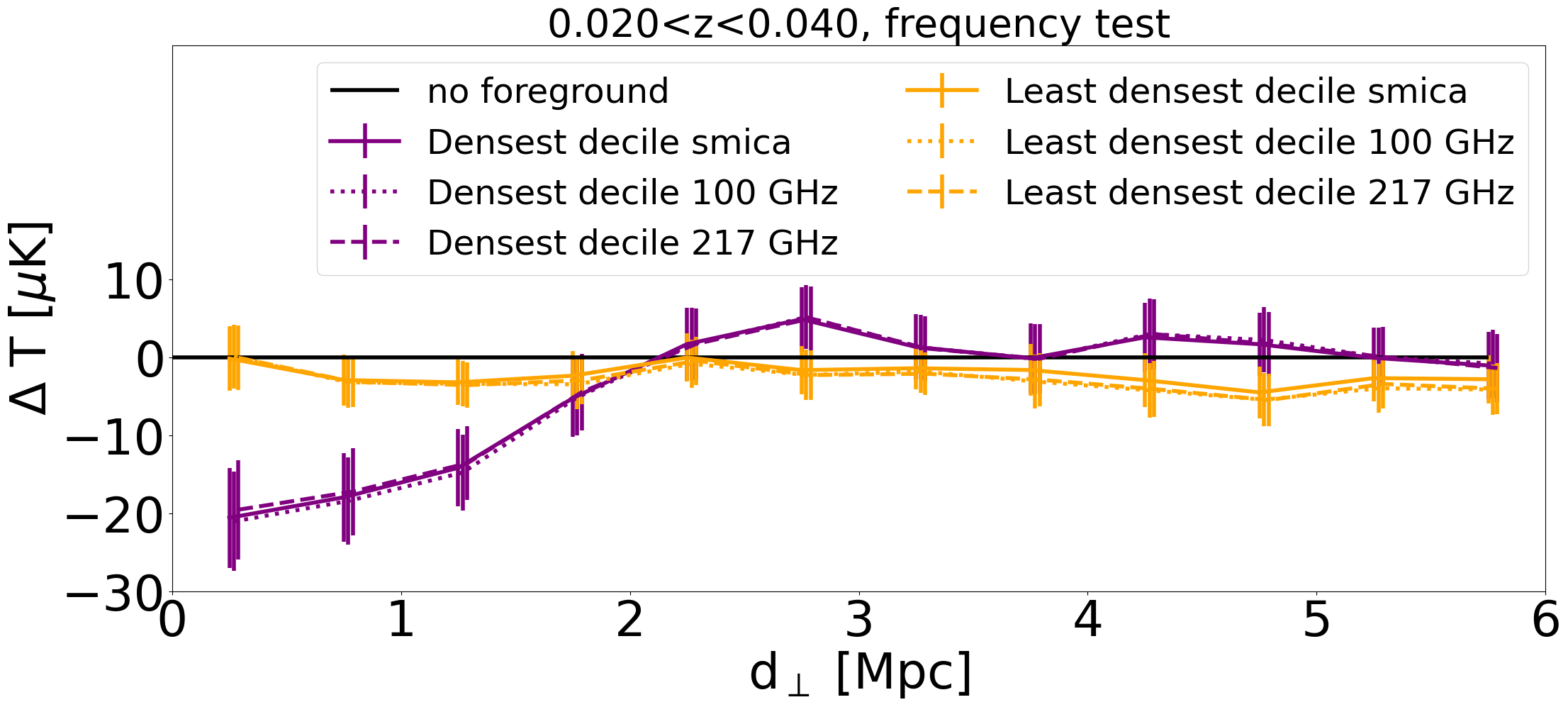}
    \caption{Same as Figures \ref{fig:fig2} (temperature profile for the densest hexile/least dense hexile for $0.004<z<0.02$) \ref{fig:fig5} (temperature profile for the densest decile/least dense decile for $0.02<z<0.04$) but now including profiles obtained on the $100$GHz and $217$GHz Planck SEVEM cleaned frequency maps.}
        \label{fig:frequencydep}
\end{figure}

In H2025 it is shown that there is no significant frequency dependence of the new extragalactic foreground associated to galaxies. A small frequency dependent CMB temperature decrement is expected for photons travelling through the filaments due to the tSZ effect discussed above. Using the full-sky map of the Compton $y$-parameter estimated from Planck frequency maps \citep{Chandran_2023}, we calculated the mean $y$-parameter and thereby the estimated CMB temperature decrement due to the SZ-effect in the vicinity of  filaments. We found that this strongly frequency dependent effect, can account at most $\sim$ 2 $\mu K$ CMB temperature decrement for the frequencies below 217GHz which we analyse, much less than our reported frequency independent signal. As the tSZ effect is zero at $217$GHz, no tSZ induced temperature decrement is expected at this frequency.

Here we confirm this frequency independence for CMB temperature profiles traverse to filaments in the two redshift ranges $0.004<z<0.020$ and $0.020<z<0.040$, using SEVEM $100$GHz and $217$GHz maps. Figure \ref{fig:frequencydep} shows the temperature profiles around the filaments within the densest/least dense hexile for $0.004<z<0.02$ and densest/least dense decile for $0.02<z<0.04$ for the SMICA maps (results presented above) as well as the $100GHz$ and $217Ghz$ SEVEM cleaned frequency maps. Note that the SEVEM cleaned frequency maps where the frequency to be cleaned has to be excluded from the cleaning process necessarily have a higher level of residual large scale galactic foreground contamination than the SMICA maps where all frequencies are used for foreground subtraction. For this reason, we do expect some deviations between the SMICA map and the SEVEM maps for larger angular scales (lower redshift filaments) as we can see in the figure where this difference is mostly seen for the least dense hexile where no signal is detected (see \cite{PlanckIV_Diffuse_component_separation} for details on the SMICA and SEVEM component separation algorithms).

Looking at the difference between the $100$Ghz and the $217$Ghz SEVEM frequency maps, there is a maximum $1-2\mu$ K difference as expected due to the tSZ effect in filaments as detailed above. Any other known scattering process of CMB photons on baryons would create a frequency dependence of the results. Also residual galactic contamination of the CMB, or a possible over-subtraction of foregrounds in Planck maps would necessarily not only be frequency depent, but also depend on the component separation method used. For a frequency independent cooling of CMB photons passing through filaments, we can therefore only see two possible but speculative explanations, either (1) in terms of an unknown CMB photon-dark matter interactions along filaments, or, (2) based on the observations in H2025b, a non-standard ISW effect (which is known to be frequency independent) associated to the filament gravitational potential evolution. Further research, both theoretical as well as observational would be needed in order to test these or other possible hypothesis.

\begin{acknowledgements}
Results in this letter are based on observations obtained with Planck (\cite{esa_planck}), an ESA science mission with instruments and contributions directly funded by ESA Member States, NASA, and Canada. We acknowledge the use of NASA’s WMAP data from the Legacy Archive for Microwave Background Data Analysis (LAMBDA), part of the High Energy Astrophysics Science Archive Center (HEASARC). The simulations were performed on resources provided by UNINETT Sigma2 - the National Infrastructure for High Performance Computing and  Data Storage in Norway". Some of the results in this letter have been derived using the HEALPix package \cite{Healpix}
\end{acknowledgements}


\bibliography{main}

\end{document}